# Alpha decay in electron environments of increasing density: from the bare nucleus to compressed matter

Fabio Belloni[*]

European Commission, Joint Research Centre, Institute for Transuranium Elements

Postfach 2340, D-76125 Karlsruhe, Germany

E-mail: fabio.belloni@ec.europa.eu

**Abstract**

The influence of the electron environment on the $\alpha$ decay is elucidated. Within the frame of a simple model based on the generalized Thomas-Fermi theory of the atom, it is shown that the increase of the electron density around the parent nucleus drives a mechanism which shortens the lifetime. Numerical results are provided for $^{144}$Nd, $^{154}$Yb and $^{210}$Po. Depending on the nuclide, fractional lifetime reduction relative to the bare nucleus is of the order of 0.1÷1% in free ions, neutral atoms and ordinary matter, but may reach up to 10% at matter densities as high as $10^4$ g/cm$^3$, in a high-$Z$ matrix. The effect induced by means of state-of-the-art compression techniques, although much smaller than previously found, would however be measurable. The extent of the effect in ultra-high-density stellar environments might become significant and would deserve further investigation.

[*] Presently at: European Commission, Directorate-General for Research & Innovation, Directorate Energy, Rue du Champ de Mars 21, 1049 Brussels, Belgium.

# 1. Introduction

Whether and to which extent the $\alpha$-decay width might be modified in electron environments has been the subject of numerous theoretical and experimental investigations, boomed over the last years. It has indeed been put forward that this effect could have implications in the stellar nucleosynthesis of heavy elements [1,2], with an impact on nuclear cosmochronology [3] (*e.g.* decay rates of $^{238}$U and $^{232}$Th [4]). Possible applications to nuclear waste management have been emphasised [3,5,6] in connection to a potentially significant increase of the decay rate in metals at low temperature. A possible exploitation of the effect for a direct measurement of the electron screening energy in nuclear reactions of astrophysical interest has also been proposed [2,7]. We add that the effect could deserve consideration in nuclear metrology, including possible implications for the accuracy of radioactive dating of geological events [8] (*e.g.* the $^{235}$U/$^{207}$Pb and $^{238}$U/$^{206}$Pb methods), as well as in geochemistry, in connection to studies of isotope cycling and fractionation on Earth (*e.g.* $^{235}$U/$^{238}$U [9] and $^{142}$Nd/$^{144}$Nd [10] ratios).

Alpha decay in a neutral atom has been addressed for the first time in the fifties [11], but convincing treatments have been provided only recently [7,12,13], which lowers the estimate of the lifetime variation relative to the bare nucleus from about 50% (early value) to a few per mil. No appreciable influence of the chemical bonding has been calculated by Rubinson and Perlman [14] already forty years ago, while a possible screening effect of free electrons in a metal lattice at low temperature [3,13] has never been measured unambiguously [15, and references therein]. By adapting Mitler's model for screened thermonuclear reactions [16], Liolios [1] has predicted a lifetime reduction by orders of magnitude in plasmas of astrophysical interest, while Zinner [17], whose calculation is based on a weak-screening perturbation to the nuclear Coulomb barrier, has come to the opposite conclusion that lifetime increases with electron density. More recently, by using a free-electron approach based on the pressure ionization model and Debye screening, Eliezer *et al.* [18] have reported that the lifetime of certain lanthanides (*e.g.* $^{144}$Nd and $^{154}$Yb) might be reduced by about 15% at low temperature, and pressures as high as those achievable in existing diamond anvil cells (DACs), ~ 4 Mbar.

In view of verifying these latter findings, and also motivated by the need for a consistent approach to the treatment of $\alpha$ decay in ions and neutral atoms, up to matter under very high pressure, we have derived a simple model for the calculation of the $\alpha$ lifetime in electron environments of different density. The basic idea underlying our study is to follow the evolution of the phenomenon starting from the bare nucleus and progressively increasing the electron density in its surroundings, encompassing this way ionization and compression within the same frame. We are convinced this approach can offer a better understanding of the phenomenon in its generality as well as an improved consistency of numerical results; as explained, previous estimates have indeed been derived over a plethora of approaches and approximations, and all together result in a non-univocal –if not contradictory– picture. The generalized Thomas-Fermi (TF) model of the atom (see *e.g.* ref. [19] for a brief review), as here specialised to the case of a finite-size nucleus, provides a reasonable compromise between accuracy and computational complexity to account for the variation of the electron density (as well as derived atomic quantities). We have then coupled the classical Gamow treatment of $\alpha$ decay in WKB approximation to the generalized TF model at $T = 0$ [†].

Here we explain, in particular, the overall mechanism ruling lifetime variation, and provide numerical results for selected nuclides, which amend and extend previous estimates. Within the approximations of our treatment, we show that the $\alpha$ lifetime in electron environments tends to decrease with increasing electron density. Directions for further theoretical and experimental work are also outlined.

---

[†] The interested reader is addressed to refs. [1,11,18,20] for previous examples of application of the TF model to screened $\alpha$-tunnelling.

**2. Theory**

*2.1. Gamow model of α decay*

In the customary *bare* model of the decay [21], the daughter nucleus and the α particle interact through a potential, $V_b$, dependent only on their relative distance $r$;

$$V_b(r) = V_s(r) + V_C(r) + V_\lambda(r)$$

where $V_s$ is the strong nuclear potential, $V_C$ is the Coulomb potential, and $V_\lambda$ is the centrifugal potential associated to the relative angular momentum λ. For the sake of simplicity, in the following we work at $\lambda = 0$, and assume $V_\lambda(r) = 0$ and

$$V_b(r) = -V_{b,0} \text{ for } r < r_0,$$

$$V_b(r) = V_C(r) = 2(Z-2)e^2/r \text{ for } r \geq r_0,$$

being $r_0$ the nuclear radius and $V_{b,0}$ the depth of the intra-nuclear potential in the square well approximation ($V_{b,0} \approx 35 \text{ MeV}$, typically; Fig. 1). In the semiclassical approximation, the decay width, $\lambda_b$, is given by

$$\lambda_b = P_b f_b \exp(-2G_b) \qquad (1)$$

where $P_b$ is the α-particle preformation probability, the Gamow factor $G_b$ is given by $G_b = \int_{r_0}^{r_1} k_b(r) dr$, and the assault frequency $f_b$ by $f_b = [(2/\mu)(V_{b,0} + Q_b)]^{1/2} / (2r_0)$, assuming a kinetic energy of $V_{b,0} + Q_b$ for the system inside the square well. The wavenumber $k_b$ is given by $k_b(r) = \{(2\mu/\eta^2)[V_b(r) - Q_b]\}^{1/2}$, where $\mu$ is the reduced mass of the system, $Q_b$ is the total kinetic energy released in the decay, and the classical turning point $r_1$ is the solution of the equation $k_b(r) = 0$. Finally, the mean lifetime $\tau_b$ is given by $\lambda_b^{-1}$.

*2.2. Extension to electron environments*

The presence of a cloud of $N$ electrons ($N \leq Z$) around the parent nucleus gives rise to a local electron density, $n(r)$, and an electron (*screening*) potential, $V_e(r)$, linked through the Poisson equation. Throughout this paper, we use the formalism where $V_e$ is taken as the interaction potential with a positive elementary charge, and is therefore a negative quantity. Here we assume that the electrons cannot penetrate into the nucleus, *i.e.* $n(r) = 0$ for $r < r_0$; accordingly, $V_e(r) \equiv V_{e,0} = V_e(r_0)$ for $r < r_0$ (Fig. 1). Generalizing the treatment for the neutral atom by Patyk *et al.* [7], the interaction potential $V_b$ and the *Q*-value are modified as

$$V_b \rightarrow V(r) = V_b(r) + 2V_e(r) \qquad (2)$$

$$Q_b \rightarrow Q = Q_b + \delta E_t \qquad (3)$$

where $\delta E_t \equiv E_{t,p} - E_{t,d} - E_{t,\alpha}$, and $E_{t,p}$, $E_{t,d}$, $E_{t,\alpha}$ are the total electron energies of the parent, daughter and He atoms (or ions), respectively.

In previous works, Eq. (3) has not been considered at all [3,11] or has been replaced with the relation $Q = Q_b + 2V_{e,0}$ [14,17,20], by analogy with Eq. (2). We remark that this latter treatment is valid in a first approximation only, the *Q*-value being always fixed by the conservation of energy between the initial and final states of the decay. As shown in Sect. 2.3, this approximation worsens with increasing electron density and systematically augments the Gamow factor, leading to the conclusion that the lifetime in an electron environment exceeds that of the bare nucleus. We actually observe that the variation of the Gamow factor with electron density is ruled by the interplay between $\delta E_t$ and $V_e$, through an interesting mechanism we now describe.

*2.3. The 'mechanism' of lifetime reduction*

When the decay occurs inside a free ion, we assume that the daughter nucleus retains all the original $N$ electrons, and that the $\alpha$ particle is emitted fully ionized (*sudden limit*); as it is obvious, the higher the ionization degree, the sounder this assumption. Accordingly, $E_t(2) = 0$ and $\delta E_t$ reads as $\delta E_t(Z, N) = E_t(Z, N) - E_t(Z-2, N)$, $N \leq Z - 2$. It is straightforward to recognize that $\delta E_t < 0$, since the ion with nuclear charge $Z$ is a more tightly bound system than the ion with nuclear charge $Z - 2$. The local electron density $n(r)$ increases with $N$ (see Sect. 3.1 and Fig. 3 for quantitative details); from basic electrostatics, $|V_e(r)|$ in turn increases with $n(r)$. Calculations as described in *e.g.* Sect. 3.1 and ref. [7] show that $V_e(r)$ varies very slowly in the tunnelling region (Fig. 1); the quantity $V_e(r_1) - V_{e,0}$ ranges from a few tens to a few hundreds eV in neutral atoms. The quantities $E_t(Z, N)$ and $V_{e,0}(Z, N)$ are linked through the Hellmann-Feynman theorem [7,22],

$$\partial E_t(Z, N)/\partial Z = V_{e,0}(Z, N)$$

which in our case yields the approximate equality $\delta E_t \approx 2 V_{e,0}$.

To summarise, we have so far shown that by *dressing* the nucleus with electrons up to the (free) neutral atom, the quantities $|\delta E_t|$ and $2|V_e(r)|_{r < r_1}$ progressively increase by similar amounts. What happens next when the electron density increases further, *e.g.* when the parent atom is embedded into a lattice in ordinary matter, up to highly compressed matter?

To this purpose, the atom can be thought as confined into a Wigner-Seitz (WS) cell [23] with radius $r_{WS}$, within which global charge neutrality holds (*i.e.*, $\int_{r_0}^{r_{WS}} n(r) d^3r = Z$); $r_{WS}$ is linked to the matter density $\rho$ through the relation

$$(4/3)\pi r_{WS}^3 \rho = A_w / N_A \qquad (4)$$

where $A_w$ is the atomic weight and $N_A$ is the Avogadro number; $\rho$ is in turn fixed by the compression factor $\eta \equiv \rho/\rho_0$, where $\rho_0$ is a reference (*e.g.* STP) density[‡]. However, one has rarely to deal with pure materials, inasmuch as radionuclides occurring in a variety of possible environments (*e.g.* stellar plasmas, naturally occurring minerals, nuclear fuel, matrices for nuclear waste conditioning) are usually diluted in complex mixtures. Even in the case of (originally) pure radioactive materials, nuclear decay unavoidably provokes a variation of chemical composition with time. In the case of mixtures, a thermodynamic equilibrium condition is required for the complete description of the electron environment (particularly the outer electron density) of each constituent element [24]. The condition of constancy of the electron chemical potential, M, is usually assumed through systems at not too large density [24,25]. Consequently, in the case of aggregate matter we will work in the simple though realistic situation of $\alpha$ emitters dispersed into a mono-elemental matrix, and will assume that

- the parent species in the initial state of the decay and the decay products in the final state are charge-neutral within their respective WS cells, and
- at rest, they hold the same chemical potential of the surrounding matrix, its value being unperturbed by effect of the decay itself.

As shown in Fig. 2a for the case of a Pb matrix, the chemical potential increases monotonically with compression. Upon compression, the total electron energy of the WS cell also increases due to the pressure-induced ionization and excitation [23,26] (Fig. 2b), vanishing or attaining even positive values at sufficiently high density. For example, from Fig. 2 one can see that $E_t$ for He becomes positive already at low compression in a Pb matrix (the reference value of $E_t$ for He is -0.078 keV at $M = 0$, *i.e.* in the isolated atom). One also notices that $|\delta E_t|$ increases upon the compression. Again, from basic electrostatics, $|V_e(r)|$ continues to increase by effect of the compression of the WS cell (Figs. 1, 2b).

---

[‡] It is noteworthy to recognize that $\eta$ also expresses the increase of the mean electron density inside the WS cell, *i.e.* $\langle n(r; \rho) \rangle / \langle n(r; \rho_0) \rangle = \eta$.

As depicted in Fig. 1, *moving* from the bare nucleus to the neutral atom (blue arrows), $Q$ shifts downward by $|\delta E_t(Z, N)|$, ranging from $Q_b$ to $Q_a$. In parallel, $V(r)$ moves downward by $2|V_e(r)|$. These shifts tend to compensate each other, so that the area of the region *ABC* (*i.e.* the Gamow factor) is negligibly affected. The net result is rather a downward translation of that region ( $ABC \rightarrow DEF$ ). Precise calculations would actually show that the Gamow factor is slightly reduced; see ref. [7] for the case of neutral atoms and hydrogenoids. Upon compression, the downward shift of $V(r)$ and $Q$ continues, but with different rates for the two quantities. Indeed, $V(r)$ decreases slightly quicker than $Q$; the net shift is represented in Fig. 2b by the quantity $\theta \equiv \delta E_t - 2V_{e,0}$. As a consequence, the area of the region *DEF* in Fig. 1 shrinks ( $DEF \rightarrow GHI$ ), which increases the penetrability of the Coulomb barrier. Physically, the behaviour of $Q$ is due to the fact that a thin fraction of the nuclear mass difference in the decay ($Q_b$) is spent in the rearrangement of the electron energy levels in the final state; accordingly, $Q < Q_b$.

## 3. Computational methods

Along the mechanism just described, we have estimated the lifetime variation in ions, neutral atoms and compressed matter for three selected nuclides (Table 1), which span representative ranges of $Z$ and half-life values, and have been considered in previous studies (*e.g.* [7,13,18]) as well. We have calculated the ratio

$$\tau/\tau_b = \lambda_b/\lambda = (f_b/f)\exp[-2(G_b - G)] \qquad (5)$$

where $\tau$, $\lambda$, $f$ and $G$ refer to the decay in the electron environment, and no change in the $\alpha$-preformation probability $P_b$ has been assumed. The quantities $f$ and $G$ are built upon the wavenumber $k(r)$, defined by analogy with $k_b$ through Eqs. (2,3). After Patyk *et al.* [7], we find convenient to write $k(r)$ in terms of the quantities $\theta$ and $\delta V_e(r) \equiv V_e(r) - V_{e,0}$;

$$k(r) = \left[(2\mu/\eta^2)|V_b(r) + 2\delta V_e(r) - Q_b - \theta|\right]^{1/2} \qquad (6)$$

We have utilized the generalized TF model of the atom to determine $V_e(r)$ and $\delta E_t$, hence $\theta$ and $\delta V_e(r)$.

While we plan to utilize more refined self-consistent field models [24,26,27] in future work, in Sects. 3.2 and 4 we show, on a phenomenological basis, that to the purpose of the calculation of $k(r)$ and within the boundaries of this study, the use of the simple TF model with the inclusion of a finite-size nucleus provides realistic approximations for $\theta$ and $\delta V_e(r)$, in spite of the fact quantum, exchange and relativistic effects are neglected.

*3.1. Application of the TF model*

In the TF treatment [19], the atomic potential $V_a(r) \equiv -Ze^2/r - V_e(r)$ is linked to $n(r)$ through the basic relation

$$n(r) = \begin{cases} (8/3)\pi(2m)^{3/2} h^{-3} [\mathrm{M} - V_a(r)]^{3/2} & \text{for } V_a(r) \leq \mathrm{M} \\ 0 & \text{elsewhere} \end{cases} \qquad (7)$$

where $m$ is the electron mass. From this, it is derived that $V_a$ can be calculated in terms of a screening function $\phi$, through the relation

$$\mathrm{M} - V_a(r) = \left(Z e^2/r\right) \phi(x) \tag{8}$$

where $x = r/\Lambda$, $\Lambda = 0.88534\, a_0 Z^{-1/3}$, and $a_0$ is the Bohr radius. The function $\phi(x)$ is the solution of the differential equation (*TF equation*)

$$\phi'' = x^{-1/2} \phi^{3/2} \tag{9}$$

with the boundary conditions $\phi(0) = 1$ and i) $\phi(\infty) = 0$ for the free neutral atom; ii) $\phi(x_q) = 0$ for ions of radius $r_q$ and ionization degree $q = 1 - N/Z$ (by the Gauss law, $q = -x_q \phi'(x_q)$); iii) $\phi(x_{WS}) - x_{WS} \phi'(x_{WS}) = 0$ for atoms in compressed matter. The chemical potential in each case can be determined via Eq. (8), calculated at the boundaries $r = \infty$, $r = r_q$, or $r = r_{WS}$, respectively, with the appropriate boundary values for $V_a$; namely, $V_a(\infty) = 0$, $V_a(r_q) = -qZe^2/r_q$, $V_a(r_{WS}) = 0$. In terms of $\phi$, $V_e(r)$ is finally calculated as

$$V_e(r) = \left(Z e^2/r\right) [\phi(x) - 1] - \mathrm{M} \tag{10}$$

Following our initial assumption $n(r) = 0$ for $r < r_0$, in the numerical integration of the TF equation we have replaced the customary point-like-nucleus boundary condition $\phi(0) = 1$ with [28]

$$\phi(x_0) - x_0 \phi'(x_0) = 1 \tag{11}$$

which expresses the constancy of $V_e(r)$ inside the nucleus and can be derived, for instance, by differentiating Eq. (10) and noticing that $V_e'(r_0) = 0$. Profiles of $\phi(x)$ generated along the boundary condition of Eq. (11) are shown in Fig. 3. We remark that the evolution of these profiles from ions to compressed atoms also describes –through Eqs. (7,8)– the corresponding local increase of $n$ (right-hand ordinate axis in Fig. 3).

The total electron energy $E_t$ has been calculated, in terms of $\phi$, as $E_t = E_k + E_p$, with the kinetic energy $E_k$ and the potential energy $E_p$ given by

$$E_k = C_k \int_{r_0}^{r_{max}} n^{5/3}(r) d^3r, \quad E_p = (1/2) \int_{r_0}^{r_{max}} n(r) [V_a(r) - Ze^2/r] d^3r$$

where $C_k = 2.1884 \times 10^{-18}$ keV·cm$^2$, $r_{max}$ stands for $r_q$, $r_{WS}$ or $\infty$, and Eqs. (7,8) are to be used. In the case of compressed matter, the electron energy calculated this way automatically includes the contribution of the free electrons, which are contained within the WS cell in the frame of the TF model; their abundance relative to the density of bound electrons is predominant in the outer layer of the cell.

The same conditions assumed by Eliezer *et al.* [18] have here been adopted for $\alpha$ emitters in compressed matter, *i.e.* nuclides dispersed into a Pb matrix ($A_w = 207.2$ g, $\rho_0 = 11.35$ g/cm$^3$). In the limit of an infinitesimal-concentration dispersion, the $r_{WS}$ value for Pb at a certain compression is straightforwardly given by Eq. (4); as a term of reference, $r_{WS} = 1.93 \times 10^{-8}$ cm at $\eta = 1$. Through Eqs. (8,9), the chemical potential of Pb is in turn set by the value of $r_{WS}$ calculated this way. As explained in Sect. 2.3, the chemical potential of Pb is then imposed to any other species hosted in the matrix, *i.e.* parent and decay-products atoms. For each species, this has been accomplished by calculating $r_{WS}$ iteratively till the chemical potential of that species equated the value for the Pb atom. In the general case of mixtures of any composition, the chemical potential has to be determined through the procedure described in ref. [24].

*3.2. Reliability considerations*

As we have verified against the collection of data for the neutral atom reported in ref. [7], solutions of the TF equation calculated upon the boundary condition $\phi(0) = 1$ yields values of $\theta$ and $\delta V_e(r)$ which are far from those obtained by more reliable relativistic quantum calculations. Consistency is recovered for $\theta$ when the boundary condition of Eq.

(11) is used; the discrepancy in $\delta V_e(r)$ also improves, even though this quantity continues to be overestimated, by a factor of about 2 in the tunnelling region. This effect is attributable to the divergence $n(r) \sim r^{-3/2}$ for $r \to 0$, which is an inherent drawback of the TF model [29] and causes the divergence of $V'_e(r)$ as well; as a consequence, values of $\delta V_e(r)$ are inflated. The impact on $\delta V_e(r)$ is obviously reduced, but not eliminated, by the use of Eq. (11), which removes the divergence but cannot change the $r^{-3/2}$ behaviour of $n$ at small $r$. The corresponding value of $k(r)$ can therefore be considered as an upper bound, the lower one being set by $\delta V_e(r) \equiv 0$ (note that $\delta V_e(r)$ is a positive-definite quantity). We extend this assumption also to the compressed atom, on the basis of the fact that –at least till $\eta \sim 10^3$– the behaviour of $\delta V_e(r)$ in the tunnelling region is highly resilient to compression. This is evident from Eq. (10) and from the behaviour of $\phi(x)$ in the tunnelling region, as opposed to that in the outer region of the WS cell (Fig. 3). Physically, this is a consequence of the fact that the core of the WS cell is nearly incompressible till the density in the outer shell has increased enough.

We also remark that, as already pointed out in ref. [7] for the case of the free neutral atom, the influence of $\delta V_e(r)$ in the calculation of the Gamow factor is generally less important than the influence of $\theta$. Nevertheless, we do note that $\delta V_e(r)$ and $\theta$ enters Eq. (6) with opposite signs ($\theta$ is also a positive quantity, Fig. 2b), which pulls the variation of the lifetime in opposite directions (increase upon $\delta V_e(r)$, decrease upon $\theta$). Critical situations may arise in the case of high-*q* ions.

Concerning the reliability of our TF calculation of $\theta$, we have already mentioned the very good agreement, for the free neutral atom, with relativistic Dirac-Fock calculations. Starting from that, after labelling by $\Delta$ variations relative to the free atom induced upon compression, we can write $\theta = \theta_a + \Delta\theta$, where $\theta_a$ is the free-atom value and, by definition of $\theta$, $\Delta\theta = \Delta(\delta E_t) - 2\Delta V_{e,0}$. The uncertainty on $\theta$ due to quantum, exchange and relativistic effects is then entirely given by the uncertainty on $\Delta\theta$. From Eq. (10), we note that $\Delta V_{e,0} \approx -\Delta M$, since the value of $\phi(r_0)$ is negligibly affected by compression (Fig. 3). We also note that $\Delta M = M$, as $M = 0$ for the free atom, and that $M$ is a positive quantity increasing with compression; see Eq. (8) and Fig. 2a. We conclude that $\Delta\theta \approx \Delta(\delta E_t) + 2M$. As seen in Sect. 2.3, the quantity $\Delta(\delta E_t)$ is positive and increases with compression; calculations show that $\Delta(\delta E_t)$ is (much) smaller than $2M$ at low/medium compression (Fig. 2), the two quantities becoming comparable only at high compression ($\eta \sim 10^3$). Reasonably, we assume that the uncertainty on $\Delta\theta$ is of the same order of magnitude than the uncertainty on $M$. As shown by Rotondo *et al.* [27], exchange corrections on $M$ are small in the low-density regime ($10^4 < r_{WS}/r_0 < 10^5$) and decrease with increasing $Z$ and compression; relativistic corrections, for elements heavier than Fe, become significant only for $r_{WS}/r_0 < 10^3$, a density regime we do not enter in our analysis. As a final remark, quantum corrections, due to the electronic shell structure, are generally expected to become negligible at high pressure [30], as the fraction of free to bound electrons increases.

## 4. Numerical results and discussion

The fractional lifetime variation $\delta\tau/\tau_b \equiv \tau/\tau_b - 1$, calculated within the limits for $k(r)$ set in Sect. 3.2, is plotted as a function of $N/Z$ and $\eta$ in Figs. 4 and 5, respectively, for selected nuclides. When shown, error bands (dashed lines) express the uncertainty on $k(r)$; mean values calculated between these upper and lower bounds are shown as continuous curves, and open circles in Fig. 4.

*4.1. Ions and neutral atoms*

Calculations for the $^{210}$Po ion are reported in Fig. 4, in the *sudden* and *adiabatic* limit of the decay, which in our formalism means $\delta E_t(Z,N) = E_t(Z,N) - E_t(Z-2,N)$ or $\delta E_t(Z,N) = E_t(Z,N) - E_t(Z-2,N-2) - E_t(2,2)$, respectively. Very different trends for $\delta\tau/\tau_b$ are obtained in the two limits; results however converge for $N/Z \to 1$, as expected. Indeed, this behaviour is determined by the difference in the values of $\delta E_t$ when calculated in the two limits; this difference increases as $N/Z$ decreases. Data in Fig. 4 also confirm our initial assumption on the sudden limit as the preferential decay mode in high-$q$ ions (Sect. 2.3); in the adiabatic limit, values of $\delta\tau/\tau_b$ at low $N/Z$ are unrealistically high and would have already been measured in experiments on highly charged ions which are currently being attempted in storage rings [2].

Mean values of $\delta\tau/\tau_b$ calculated in the adiabatic limit at $N/Z = 1$ are reported as $\delta\lambda/\lambda_b$ in Table 1, for the three selected $\alpha$ emitters. These values indicate a slight reduction ($\delta 1\%$) of the lifetime and are highly compatible with the neutral-atom calculations of Patyk *et al.* [7], based on relativistic electron binding energies.

Sudden-limit datapoints in Fig. 4 show oscillations which are probably due to limited accuracy in the prediction of $\theta$. The amplitude of these oscillations tends to increase at low $N/Z$, and it is indeed well known that the TF model is capable to provide a reasonable estimate of energies and potentials in ions only in the large-$Z$, large-$N$ limit [31]. We plan to undertake further work in order to verify the consistency of these data, in the limit $\delta V_e(r) = 0$, by using the precise electron binding energies of the isoelectronic series of Rodrigues *et al.* [32], and values of $V_{e,0}$ calculated by means of the Hellmann-Feynman theorem as in ref. [7].

Sudden-limit data generally show a reduction of lifetime in ions compared to the bare nucleus. Variation is constrained within 0.5% for $^{210}$Po, and is compatible with the value for the H-like ion calculated (at $\delta V_e(r) = 0$) by using relativistic electron binding energies, as described in ref. [7]. A straight-line fit to the sudden-limit data is also shown in Fig. 4; again, its value at $N/Z = 1$ is in good agreement with the adiabatic-limit prediction.

*4.2. Compressed matter*

Curves in Fig. 5 show a stronger reduction of lifetime. Typical uncertainty due to $\delta V_e(r)$ is shown for $^{154}$Yb. At high compression, $\theta$ increasingly leads over $\delta V_e(r)$ in Eq. (6), reducing this uncertainty significantly.

Unlike Eliezer *et al.* [18], we do not observe a special effect on $^{144}$Nd and $^{154}$Yb, compared to $^{210}$Po, in the region $1 < \eta < 10$. Fractional lifetime variation is of the same order of magnitude for $^{154}$Yb and $^{210}$Po (between 0.1% and 1%). Effects on $^{144}$Nd are somewhat larger (between 1% and 2%). Lifetime variation proceeds slowly in this region; while $\tau$ is nearly stable for $^{154}$Yb, it slightly increases (meaning a decrease in $-\delta\tau/\tau_b$) in the case of $^{210}$Po and $^{144}$Nd.

In a possible experiment on compressed matter, values of $\tau$ measured at high pressure would be compared with the reference value, $\tau_0$, measured in matter in ordinary conditions ($\eta = 1$). In terms of the quantity $\delta\tau/\tau_b$ we have calculated, one finds straightforwardly

$$\frac{\delta\tau}{\tau_0}(\eta) = \left[\frac{\delta\tau}{\tau_b}(\eta) - \frac{\delta\tau}{\tau_b}(1)\right] \cdot \left[1 + \frac{\delta\tau}{\tau_b}(1)\right]^{-1} \approx \frac{\delta\tau}{\tau_b}(\eta) - \frac{\delta\tau}{\tau_b}(1)$$

Our data for $^{210}$Po and $^{144}$Nd show that $\delta\tau/\tau_0 \approx 0.1\%$ at the highest static compressions achievable in the lab (*e.g.* $\eta \approx 2.3$ on Pb [33]). Although much smaller than previously found by Eliezer *et al.* [18], the effect of compression could however be still measurable, by using suitable nuclides and state-of-the-art high-pressure techniques (notably, DAC compression), as originally proposed by those authors. Effects of this magnitude or even smaller have already been measured in high-pressure experiments on *e.g.* $^{7}$Be (electron capture) [8] and $^{99m}$Tc (isomeric transition) [34]. In these experiments, the activity of the sample has been monitored over time through the detection of $\gamma$ rays associated to the decay. Similarly, the detection of $\gamma$ (or X) rays emitted by the daughter nucleus (atom), or by nuclides of the decay chain in secular equilibrium with the parent, could be used in an experiment on $\alpha$ decay.

Due to the tiny lifetime variations we are confronted with, the improvement of our estimates of $\delta\tau/\tau_0$ in the low-$\eta$ region should account for quantum and relativistic effects in the calculation of $\theta$, $\delta V_e(r)$ and M, and would need a further modelling (as well as computational) effort. The adoption of the relativistic Hartree-Fock-Slater self-consistent field model of Rozsnyai [26] would be appropriate in this respect. In addition, a more realistic nuclear potential should be adopted for $V_b(r)$, *e.g.* that used by Buck *et al.* [21] for systematic calculations of $\alpha$ half-lives.

Above $\eta \approx 10$, the trend of the curves in Fig. 5 becomes more sloping, although saturation appears for $^{144}$Nd and $^{154}$Yb around $\eta = 1000$. Lifetime reduction ranges from approximately 3% to 10% –depending on the nuclide– at $\eta = 1000$, a density regime beyond which relativistic effects can no longer be neglected. In this connection, we remark the behaviour of the pre-exponential factor $f_b/f$ in Eq. (1). In fact, one easily finds $f_b/f \approx 1 - (1/2)\theta/(Q_b + V_{b,0})$. This factor therefore tends to depress the ratio $\tau/\tau_b$, although we find its effect negligible ($1 - f_b/f < 10^{-4}$) in the domain of matter density considered here. Its impact could however become of some importance in the relativistic domain of the equation of state, where an increasing trend can be extrapolated for $\theta$ (Fig. 2b).

In our calculations, lifetime variation is therefore solely attributable to the perturbation of the Gamow factor. We note, however, that other forms are possible for the expression of the pre-exponential factor in Eq. (1), which imply a stronger dependency on $Q$. The assault frequency is often taken to be proportional to $\sqrt{Q}$ [35], meaning that the kinetic energy of the daughter/$\alpha$ system inside the potential well is of the same order of magnitude of the kinetic energy at infinity. Based on the phenomenology of the zero-point vibration energy, other authors [36] have utilised $f \propto Q$ in systematic calculations of $\alpha$ half-lives. It has also been shown [13] that for $\alpha$ channels far from the resonance of a quasistationary state inside the potential well, the pre-exponential factor scales as $\sim Q$. Calculations of $\delta\tau/\tau_b$ performed with a pre-exponential dependency on $Q$ stronger than that used in this work would therefore lead to somewhat different numerical results. For example, a pre-exponential factor of the form $\sim Q^\sigma$, $\sigma = 1/2, 1$, would result in a shift of $\delta\tau/\tau_b$ toward positive values, by the quantity $\sim -\sigma\, \delta E_t/Q_b$, where $-\delta E_t/Q_b$ is of the order of 1%, typically [13]. On the other hand, precise measurements of lifetime variation between bare nucleus and neutral atom could give indications on the actual form of the pre-exponential factor.

*4.3. Comparison with previous works*

We have already commented about the findings of Patyk *et al.* [7] and Eliezer *et al.* [18] in Sects. 4.1 and 4.2, respectively. Here we remark that our numerical results reconnect to those of Zinner [17], Igashov and Tchuvil'sky [20], Karpeshin [12], and Dzyublik [13] in the limit $\theta = 0$ (*i.e.*, $\delta E_t = 2V_{e,0}$; see Sect. 2.2). All these authors find positive values of $\delta\tau/\tau_b$. In particular, the mechanism of lifetime variation outlined by Zinner, which is qualitatively similar to that we describe in Sect. 2.3 (the quantitative difference being played by $\theta$) results in the suppression of the decay (*i.e.* infinite lifetime) when extrapolated to the limit of ultra-high-density stellar environments. On the contrary, reasoning and calculations presented in this article point toward an increased instability of $\alpha$ emitters in those environments, in qualitative agreement with previous findings by Liolios [1]. Nevertheless, an extrapolation of our results appear to be orders of magnitude lower than the predictions of that author. Moreover, we notice that the extrapolation of the curves of $^{144}$Nd and $^{154}$Yb in Fig. 4 is uncertain, since some evidence of saturation appears. We also notice that our formalism reconnects to that used by Liolios in the limit $\delta V_e(r) = 0$ and $\delta E_t = 0$, $2V_{e,0} = -U_e^M$ (implying $\theta = U_e^M$), where $U_e^M$ is the Mitler's shift [16] as modified in ref. [1]. We plan to verify our results at high $\eta$ and extend them to the relativistic domain of the equation of state in future work, by adopting the relativistic Thomas-Fermi-Dirac statistical treatment of the compressed atom described in ref. [27].

Finally, we comment about a possible role of temperature, also considered in previous studies on both the theoretical (*e.g.* [3,17,18]) and experimental grounds (*e.g.* [6,15]). From our treatment, it straightforwardly emerges that the influence of temperature is certainly negligible in ordinary matter, and is likely to represent a second-order effect in stellar environments. As a matter of fact, it is well known that the TF treatment in the $T = 0$ limit can be retained up to $kT$ of the order of the chemical potential [24]. We find $kT \ll M$ for STP matter, and notice that, although $kT$ may be extremely high in stellar plasmas, M is also increasingly large in those dense environments.

*4.4. Extension to matrices of other elements and implications for nuclear metrology*

As we have verified, differences in the values of the chemical potential calculated in the Pb matrix for $\eta < 3$ and those calculated in pure matrices of other elements at the same $\eta$ (referred to $\rho_0$ at STP in each case) are within a few eV. These differences increase with $\eta$ and become significant in the high-$\eta$ region. On the other hand, the difference in the values of M between $\eta = 3$ and $\eta = 1$ for Pb is about 20 eV. As seen in Sect. 4.2, this corresponds to a lifetime variation up to 0.1% in certain nuclides. Since in our model the effect of the matrix on the decay is exclusively parameterized in terms of M, we conclude that differences in the values of $\delta\tau/\tau_b$ calculated at a given $\eta$ in different matrices (*i.e.*, the fractional half-life difference between two matrices) must be much smaller than $10^{-3}$, for $\eta < 3$. This conclusion is in line with precise half-life measurements on $^{221}$Fr implanted into matrices of Si and Au at ordinary density [37].

Previous considerations imply that the results given for the Pb matrix in the low-$\eta$ region are practically matrix-independent and can be considered of general validity for (mono-elemental) aggregate matter.

## 5. Conclusion

We have addressed the controversial problem of $\alpha$-lifetime perturbation by the effect of electrons in the nuclear surroundings. Following previous work by Erma [11], Patyk *et al.* [7], and Eliezer *et al.* [18], we have derived a simple model suitable to the calculation of the decay width in electron environments of increasing density, ranging from ions to the neutral atom, up to highly compressed matter. A unitary approach has been achieved by coupling the customary Gamow model of $\alpha$ decay to the generalized TF theory of the atom. Our reasoning and calculations set a coherent picture along which the perturbed $\alpha$-lifetime generally decreases while *moving* from the bare nucleus to high-density matter. Numerical results for selected nuclides indicate lifetime variations of the order of 0.1÷1% in free ions, free neutral atoms and ordinary matter. The extent of the effect may reach up to 10% at a density of the order of $10^4$ g/cm$^3$, in a high-$Z$ matrix. Where appropriate, comparison has been made with earlier findings as well.

We have also shown that $\alpha$ decay occurs in the sudden limit in highly charged ions, and that a thermodynamic equilibrium condition for parent, daughter and He atoms in the host matrix is needed to ensure a consistent description of the decay in compressed matter. The equilibrium condition here adopted is the equality of the chemical potential, which in turn depends not only on compression, but also on chemical composition of the matrix. We do remark, however, that the electron pressure at the WS cell boundary is a more general equilibrium parameter and its use is more appropriate at very high compression. The strong implication we here wish to put forward, is that the decay process is influenced by the broader environment of the matrix where it takes place, and that a correct prediction of the decay width –especially at high compression– can not be separated from the equation of state of the surrounding matter.

Experiments by means of state-of-the-art compression techniques are encouraged, which would allow to validate our findings and could detect the first clear evidence of lifetime variation of an $\alpha$ emitter induced by its physical environment. Finally, we point out the need for further studies –based on more refined self-consistent field models– addressing the implications of this effect for synthesis and abundance evolution of $\alpha$-decaying nuclides in stellar plasmas.

**Acknowledgements**

Stimulating discussions with R. Caciuffo, T. Fanghänel, F. Raiola and D. Delle Side are gratefully acknowledged. The author is indebted to F. Wastin, J.-P. Glatz and G. Tamborini for reviewing this work prior to its submission for publication. Finally, the valuable support of D. Ostojič is gratefully acknowledged.

# Tables

**Table 1.** Relevant data for $0^+ \to 0^+$ $\alpha$ decays selected for calculations.

| Parent | Daughter | $BR$[a,b] (%) | $T_{1/2,\alpha}$[a,c] (s) | $r_n$[a] (fm) | $Q_a$[a,d] (MeV) | $Q_b$ (MeV) | $\delta\lambda/\lambda_b$[e] (%) |
|---|---|---|---|---|---|---|---|
| $^{144}_{60}$Nd | $^{140}_{58}$Ce | 100.0 | 7.2 x $10^{22}$ | 6.712 | 1.902 | 1.924 | 1.0 |
| $^{154}_{70}$Yb | $^{150}_{68}$Er | 93.0 | 0.44 | 6.758 | 5.484 | 5.511 | 0.3 |
| $^{210}_{84}$Po | $^{206}_{82}$Pb | 100.0 | 1.2 x $10^7$ | 6.966 | 5.439 | 5.473 | 0.2 |

[a] From Buck *et al.* [21].

[b] Branching ratio for $\alpha$ decays that populate the $0^+$ daughter state (given as percentage of all decays).

[c] Measured partial $\alpha$ half-life, *i.e.* observed half-life of the given nuclide divided by *BR*.

[d] *Q*-value deduced from the measured $\alpha$-particle kinetic energy; here it is assimilated to the value for the free atom, and used in Eq. (2) to calculate $Q_b$.

[e] Value calculated for the free neutral atom.

**Figure captions**

Figure 1. (Color online) Nuclear and screening potentials in $\alpha$-decay tunnelling, as a function of the daughter-nucleus/$\alpha$-particle relative distance. The evolution of potentials, $Q$-value and Gamow factor with increasing electron density is shown, from the bare nucleus to the free neutral atom (blue arrows), to highly compressed atoms (red arrows). Data refer to $^{210}$Po; separation among the $Q$-values and among the nuclear potential curves is magnified by a factor of 100.

Figure 2. a) Chemical potential for Pb as a function of compression. Data are calculated through the variation of $r_{WS}$ (Pb-nat. values $A_w = 207.2$ g, $\rho_0 = 11.35$ g/cm$^3$ are used), as explained in Sect. 3.1. b) Evolution of $E_t$, $\delta E_t$, $V_{e,0}$ and $\theta$ as a function of chemical potential for the species involved in the $\alpha$ decay of $^{210}$Po. The chemical potential is calculated independently for each species by varying $r_{WS}$. A curve like the one in panel a) can be used to relate plotted quantities to compression of the host matrix, the (common) value of M being fixed by the matrix.

Figure 3. Evolution upon ionization and compression of the profiles of the TF screening function, calculated upon the boundary condition of Eq. (11).

Figure 4. (Color online) Fractional lifetime variation relative to the bare nucleus, as a function of $N/Z$, for the $^{210}$Po free ion.

Figure 5. (Color online) Fractional lifetime variation relative to the bare nucleus, as a function of compression, for selected nuclides dispersed into a Pb matrix ($\eta = 1$ at $\rho = 11.35$ g/cm$^3$).

Relevant physical environments and compression techniques are also indicated as terms of reference for $\eta$ or $\rho$ (e.d.: *equivalent density*; denotes an environment with the same density of the Pb matrix at the value of $\eta$ indicated).

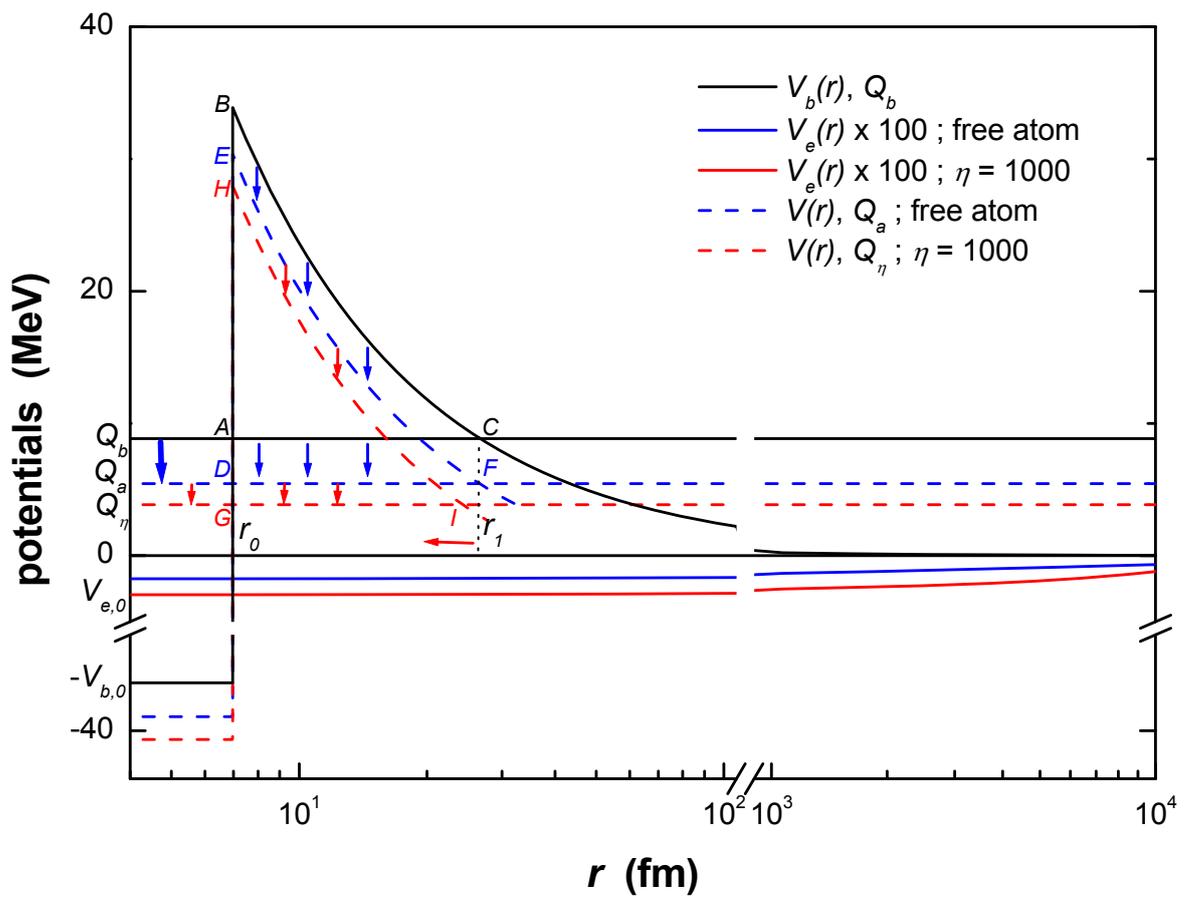

Fig. 1 – F. Belloni

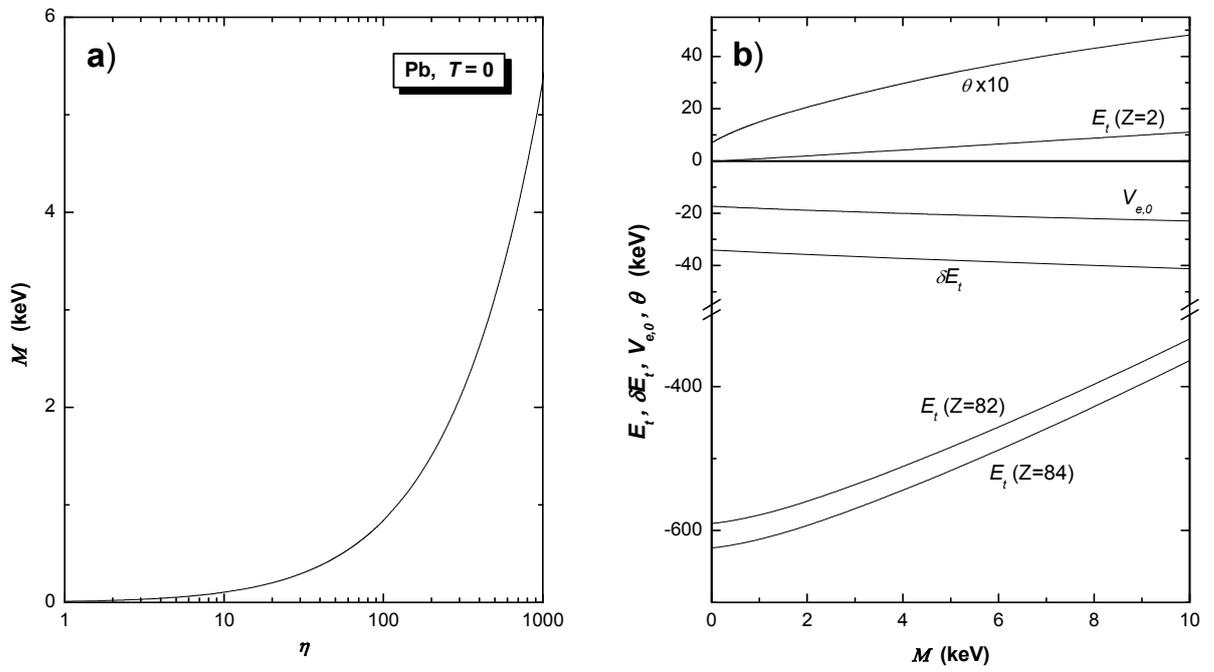

Fig. 2 – F. Belloni

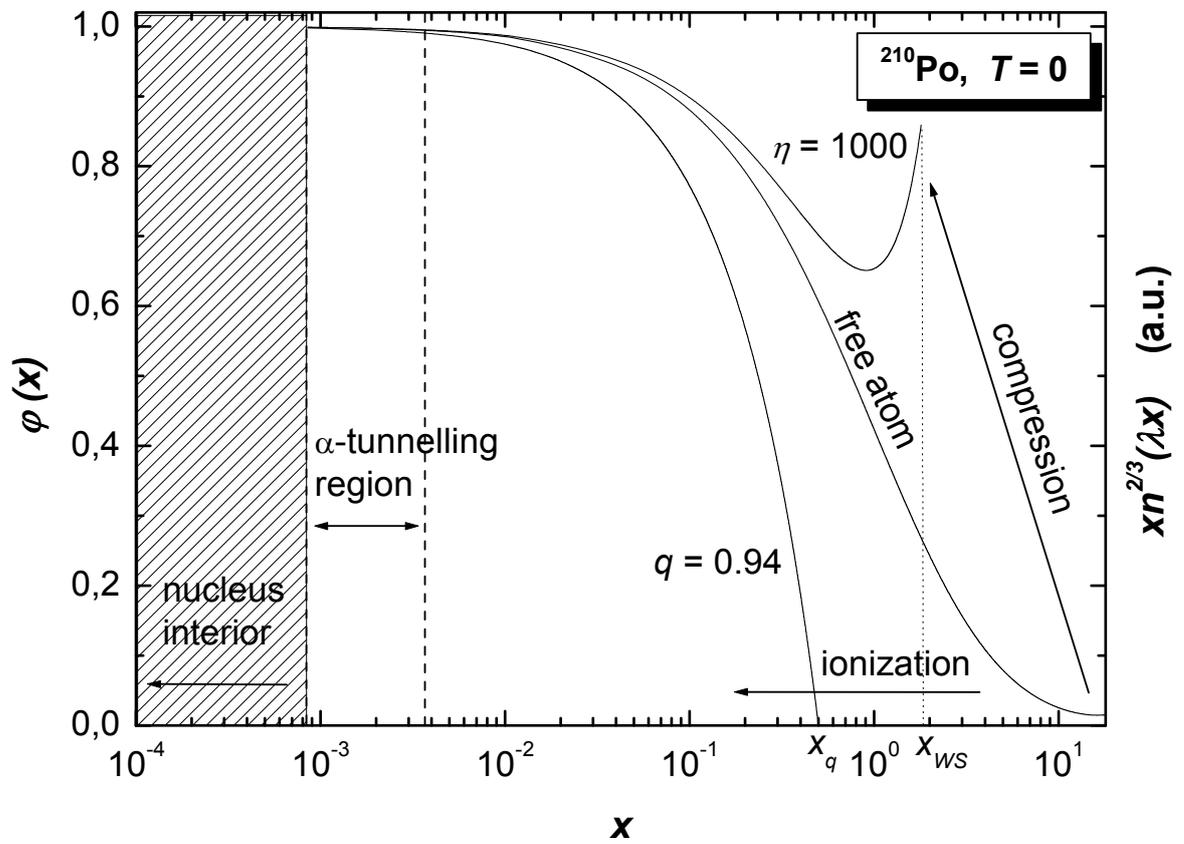

Fig. 3 – F. Belloni

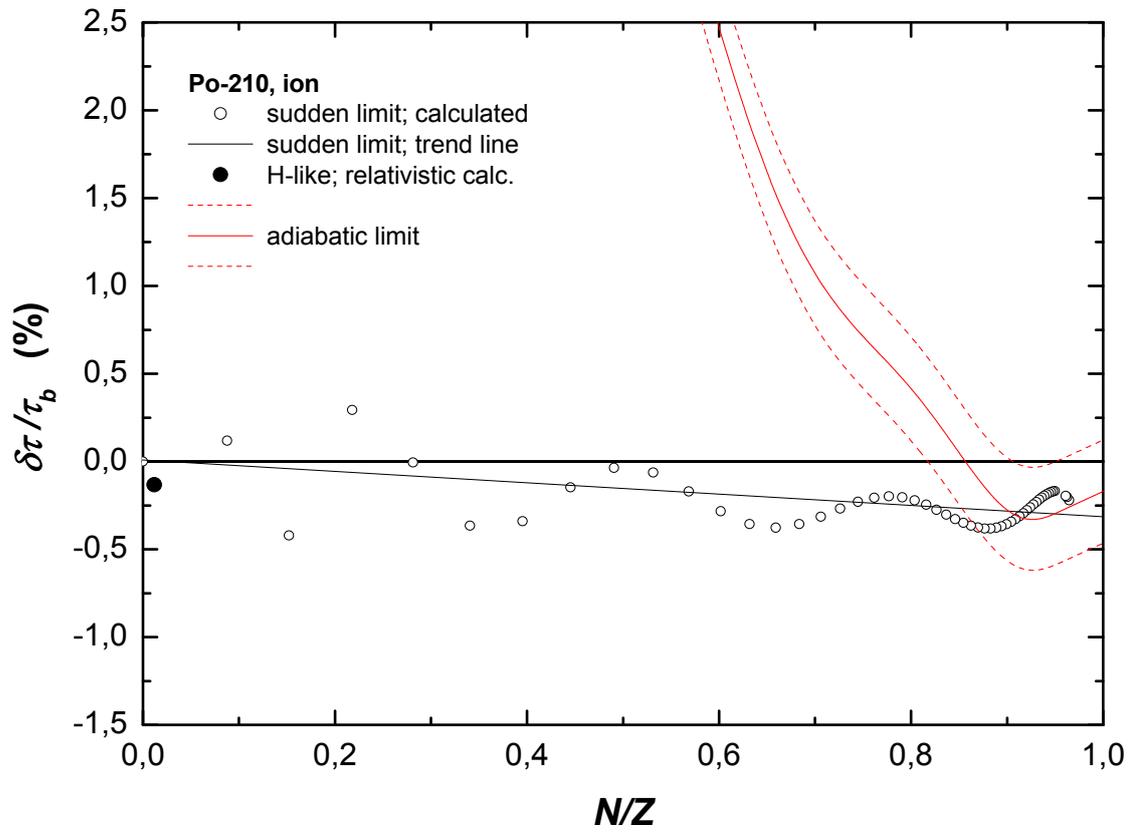

Fig. 4 – F. Belloni

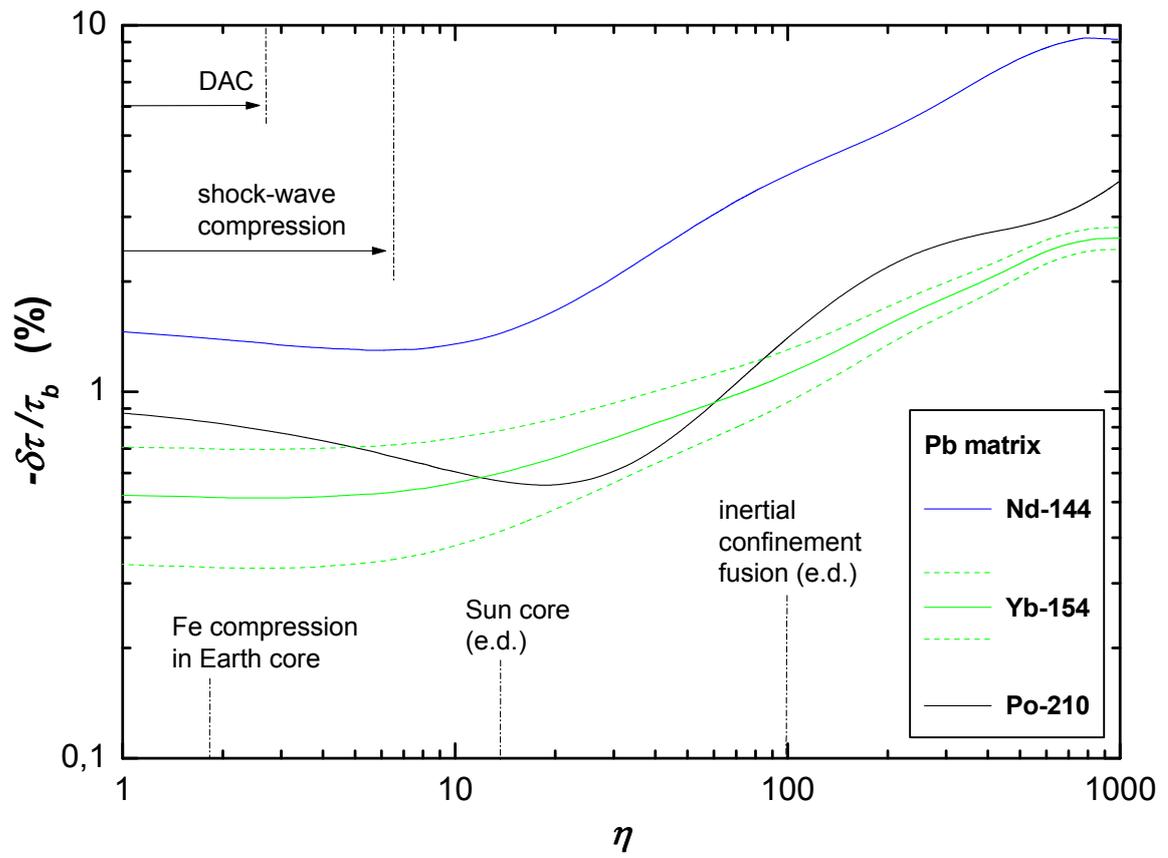

Fig. 5 – F. Belloni